\documentclass[12pt,showpacs,showkeys,nofootinbib]{revtex4}
\usepackage{graphicx}
\usepackage{psfrag}
\usepackage{amsmath}
\usepackage{amssymb}
\usepackage{bm}
\usepackage{bbm}
\begin{document}

\title{Positronium decay in a circular polarized laser field}

\author{Yu-Qi Chen $^{1}$}
\address{$^1$Key Laboratory of Frontiers in Theoretical Physics,
\\The Institute of Theoretical Physics,  Chinese Academy of Sciences,
Beijing 100190,  People's Republic of China\vspace{0.2cm}}
\author{Pei Wu $^{1}$}

\address{$^1$Key
Laboratory of Frontiers in Theoretical Physics,
\\The Institute of Theoretical Physics,  Chinese Academy of Sciences,
Beijing 100190,  People's Republic of China\vspace{0.2cm}}



\begin{abstract}

We calculate the lifetime of both the o-Ps and the p-Ps positronium
annihilation decay $ Ps\to\gamma\gamma$ in the strong circular
polarized laser field. We take a strategy of the factorization to
separate the effects caused by the Coulomb interaction and the
strong laser field interaction. It is factorized in  the time
direction but not in the space direction. Our results show that in
the laser  with long wavelength and high intensity, the lifetimes of
those Ps states are dramatically increased. For $\rm CO_2$ laser
with $10\, \mu m$ wavelength and  $10^{13} W/cm^2$ intensity,
lifetime of the spin-single positronium is increased by $10^8$
times. Our result is consistent with those obtained by solving the
Sch{\"{o}}dinger equation. This effect may be useful for the high
harmonic generation(HHG) effects provided with the
Ps\cite{keitel2004}.

\end{abstract}



\maketitle
\newpage


\section{Introduction}
\label{introduction}


The fast technological advance in the past decades brings us the
high-power laser  systems  with  peak intensities up to $10^{22}
W/cm^2$\cite{laserintensity22}, and one would expect a further
increase in the near future\cite{275004}.
Such high-power laser systems lead to studies of the fundamental
processes of quantum electrodynamics in the presence of a strong
laser field realistic in experiments. Some typical processes  are
laser-induced Comptom scattering\cite{compton}, laser-assisted Mott
scattering\cite{Mott}, laser-assisted M$\phi$ller scattering
\cite{Moller1, moller2}and laser-assisted Bhabha
scattering\cite{Bhabha}, bremsstrahlung\cite{bremsstrahlung}. The
mechanism of production of $e^+e^-$ pair creation by a projectile
particle colliding with an intense laser beam is widely studied both
theoretical and experimental, and first directly observed at SLAC
facility\cite{46.6}.

Positronium(Ps), the electron-positron bound state under attractive
Coulomb interactions, is well suitable for probing many fundamental
aspects of particle physics. For the Ps in vacuum, there are three
distinct energy scales: the electron mass $m$, the three-momentum
$m\alpha$ and the binding energy $m\alpha^2$. The lowest $S-$wave
state can be spin-singlet para-positronium(p-Ps) or spin-triplet
ortho-positronium(o-Ps). The binding energy of these two states is
$1/2\, m \,\alpha^2 = 6.8 eV$. They can decay into even or odd hard
photons, respectively, via spontaneous annihilation. The spin 0
singlet state annihilates into two photons with the lifetime of
$1.25\times10^{-10}$ second while the spin one triplet decay only
into three photons with a lifetime of $1.4\times10^{-7} $ second.

Under strong laser fields, besides the annihilation decay, the  Ps
may also decay via ionization. The ionization of normal atoms under
laser is described by the Keldysh's theory\cite{Keldysh}. In laser
electro-magnetic (EM)  field, the charged particles receive the
ponderomotive motion along the laser beam direction which is a
nonlinear effects of the EM interaction of the electrons and the
laser beam. For normal atoms, they are ionized since the nucleus and
the electrons gain very different velocities because their masses
are very different. However, for the Ps, the ionization is quite
different from that of the normal atoms for the reason that both the
electron and the positron in the Ps receive the same ponderomotive
momentum along the laser beam direction since they possess the same
mass. Consequently, the Ps is much harder to be ionized than the
normal atoms under the high laser EM field. This feature is
suggested to produce HHG by the Ps state\cite{keitel2004}. To make
this mechanism work, it is required that this has enough lifetime.
Thus it is essential to calculate the partial lifetime of the
annihilation process.


Under strong laser EM fields, the annihilation partial decay width
of the Ps may be changed. Intuitively,  with the laser field, the
electron and the positron move back and forward periodically in the
perpendicular direction of the laser beam. It effectively reduces
the wavefunction at origin and enhances the lifetime of the Ps.
Unlike in vacuum, in laser EM fields, both the o-Ps and the p-Ps can
decay into two gamma photons. Meanwhile the Ps annihilation may
produce a single photon. To precisely predict the lifetime of the
Ps, one needs to solve the dynamics of the system combining Coulomb
potential, the laser potential and the annihilation process
together, which is very complicated. To carry out calculations,
certain approximations are necessary.



In Refs.\cite{Ehlotzky}\cite{Brazil}, the authors studied the
nonperturbative effects of intense laser field on the origin of the
wavefunction by solving Sch{\"{o}}dinger equation (SCHE). They
started from a time-dependent SCHE with Coulomb interaction and the
dipole interaction for the laser EM field which is reasonable since
the wavelength of the laser photon is much larger than the typical
size of the Ps. With some additional approximations they derive a
time-independent local  SCHE with the dynamics similar to the $\rm
H_2^+$ ion. By using variational method,  they can approximately
solved the equation. Their method is available only for the linearly
polarized laser field.



In this paper we reconsider the Ps decay processes into two photons
in circular polarized laser field with higher intensity by examining
the energy scales involved in the annihilation decay processes.
Notice that  besides  two produced hard photons, there are a set of
laser photons are emitted or absorbed from the laser sources. Thus
the basic annihilation processes can be described as:
\begin{equation}
e^+ + e^- \to \gamma + \gamma \pm n \, \gamma_k \;, \label{epem}
\end{equation}
where  $\pm n $ is the number of emitted or absorbed laser photon
$\gamma_k$ with  momentum  $k = (\omega, {\rm\bf k})  $. When the
typical $n$ satisfying $n\omega $ much larger than the binding
energy of the bound state,  the typical kinetic energy of the
relative motion in the rest frame of the $e^+ e^-$ system is then
much larger than the Coulomb potential energy. Hence the Coulomb
potential interaction can be negligible in the leading order
approximation in calculating this annihilation process. In this way,
the effects of the annihilation decay process (\ref{epem}), as a
short-distance effect, is factored out from the long-distance
effects which is governed by Coulomb interaction. This factorization
is different from the conventional factorization. It is factorized
in the time direction but not in the space direction since they are
in the compatible  size in the space direction.

The short-distance effect can be calculated by perturbation theory.
It is different from the usual perturbation theory in vacuum. Here
one has to include contributions arising from this background EM
field. To this end, we take Volkov state to describe the electron
and the positron under the classic laser EM field\cite{Volkov}. To
carry our calculations, we also use the propagator of the electron
and the positron under the laser EM background field obtained by
Schwinger proper-time method. Our strategy is in the spirit of  the
NRQED factorization formula, which is the QED version of NRQCD
factorization formula\cite{BBL} for strong interaction without laser
field background. Our results are suitable for the case that the
motions of the positron and the electron caused by the laser field
is relativistic. Moreover, one advantage of our method is that it
can be used to calculate various differential distributions of the
final hard photons. In addition, it can be applied to the case of
circular polarization laser.

The remainder of this paper is organized as follows. In Sec.~II, we
propose a factorization formula to carry out  calculations on the
processes. In Sec.~III, we present numerical results.  Sec.IV is
dedicated to a brief summary and conclusions. Throughout this paper,
we use natural units with $\hbar=c=1$. The fine structure constant
is $\alpha=\frac{e^2}{4\pi}\approx \frac{1}{137}$.


\section{Theoretical Framework}
\label{formulation}

In this section, we calculate the decay rate of the Ps into double
hard photons in a strong laser EM field.

\begin{figure}[ht]
\begin{center}
\includegraphics*[scale=0.8]{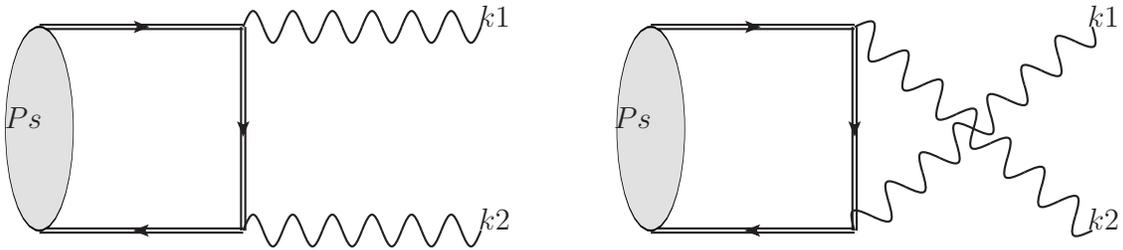}
\caption{Feynman diagrams for $Ps \to \gamma\gamma$ in a laser
field.Double lines correspond to the electron(positron) Volkov state
} \label{feynman:fig}
\end{center}
\end{figure}

The process can proceed at the second order in the QED hard
interaction  with momentum transfer at order of $m$ or higher. There
are  two Feynman diagrams contributing to the process as shown in
Fig.~\ref{feynman:fig}. For the photons with momenta $k_1$ and $k_2$
and corresponding polarization vectors
$\mbox{\boldmath$\epsilon$}_1$ and $\mbox{\boldmath$\epsilon$}_2$,
respectively,  the S-matrix elements can be written as
\begin{eqnarray}
S_{Ps \to \gamma\gamma} &=& -e^2\;
 {\epsilon_{1 \mu}   \epsilon_{2 \nu} }
 \; \int d^4x_1\, d^4x_2
 \;  e^{ i\,  k_1  x_1 + i\,k_2 x_2   } \nonumber \\
& \times& 
     \langle 0 | \,\left[\, \bar{\Psi}\left( {x_1  } \right)
     \, \gamma^\mu \,S \left( {x_1  }, {x_2  } \right) \,
           \gamma^\nu
    \Psi \left(  {x_2 } \right)
 \, + \,
      \bar{\Psi}\left( {x_2  } \right)
     \, \gamma^\nu \,S \left( {x_2  }, {x_1  } \right) \,
           \gamma^\mu
     \Psi \left(  {x_1  } \right) \,\right]\,| Ps \rangle
  \;,
\label{S-Psgg}
\end{eqnarray}
where $\Psi(x)$ and  $S \left( {x_1  }, {x_2  } \right)$ are the
four-fermion field and the propagator of the electrons under the
Coulomb interaction and the laser EM field. They satisfy Dirac
Equation
\begin{equation}
( i\not \!\! {D} -m )_x \; \Psi(x) = 0 \;, \label{dirac}
\end{equation}
and
\begin{equation}
( i\not \!\! {D} -m )_x \; S (x,y) = \delta^4 (x-y) \;,
\label{pro-elec}
\end{equation}
respectively, where $ D_\mu = \partial _\mu + i e A_\mu $. Here the
vector  potential $A_\mu$ consists of Coulomb potential  and the
laser EM potential. The laser EM potential appearing here implies
that  arbitrary number laser photons can be absorbed or emitted by
the electron and the positron. Generally, these equations are
difficult to be solved analytically. Here our strategy is to
disentangle the long distance effects and the short distance
effects. Notice that in the case of the laser field absent, the
annihilation of the electron and the positron is the short-distance
effect while the formation of the bound state is the long-distance
effect. They are separated well and factorizable. When the laser
field appear, the fermion line will emit or absorb a number of laser
photons from the laser source. Including these photons, the real
annihilation process is the one given in Eq.~(\ref{epem}). The
stronger of the laser field is, the larger  the typical $n$ is. If
the laser field is sufficiently strong so that  $n \omega$ is much
larger than the binding energy of the positronium, the electron or
the positron gain typical kinetic energy (either with plus sign or
with minus sign) of the relative motion in the rest frame of the
$e^+ e^-$ system being much larger than the Coulomb potential
energy. Hence the contribution from the Coulomb potential
interaction can be negligible in calculating the interaction field
by Eq.~(\ref{dirac}) and the propagator by Eq.~(\ref{pro-elec})
appearing in Eq.~(\ref{S-Psgg}). In the diagrammatic language, the
bound state of the Ps is formed by exchanging Coulomb interactions.
One has to resum over all those diagrams since each of them give the
same order contribution to the binding energy. Specifically, the
Coulomb singularity contribution ($1/v \sim 1/\alpha $ ) arising
from each Coulomb photon exchange is compensated by the coupling
constant $\alpha$ from one more exchange. Once $n \omega$ is much
larger than the binding energy of the positronium, the electron or
the positron are far away from the Coulomb region with momentum $(
m\alpha^2, m\alpha )$, the Coulomb singularity appears no longer
while the coupling is still there. Thus one more such exchange of
the ladder is suppressed by a factor of $\alpha$, hence
perturbatively calculable. In the leading order approximation, one
may just neglect it. Therefore, in solving the Dirac
Eq.~(\ref{dirac}) and the propagator by Eq.~(\ref{pro-elec}), one
may neglect the Coulomb interactions once the typical value of $n$
satisfies $n\omega \gg m \alpha^2 $ and then the calculation can be
simplified significantly. This is just the case we are concerning in
this paper. In this way,  the long-distance effects caused by the
Coulomb interaction in the annihilation process are factored out.
However, this factorization is different from the conventional
factorization. It is factorized in the time direction but not in the
space direction since they are in the same size in the space
direction. With only the laser EM field, those equations can readily
be solved.


We assume the laser  propagates along the $z$ direction with photon
momentum  $k^\mu=\omega(1, 0, 0, 1)$ and the vector potential $A$
possesses only  transverse components. For a circular polarized
laser, the vector potential can be written as:
\begin{eqnarray}
A^{\mu}(x)=a^\mu_1\cos\xi+a^\mu_2\sin\xi \,
\end{eqnarray}
where $a^\mu_1=a(0, 1, 0, 0)$, $a^\mu_2=a(0, 0, 1, 0)$ , $a_1\cdot
a_2=0$, $\xi=k\cdot x$. Thus $a$ denotes the amplitude of the vector
potential. For the  circular polarized laser, one has $A^2=-a^2$.

One useful dimensionless parameter describing nonlinear effects is
$\eta $ which is defined as:
\begin{eqnarray}
\eta \equiv \frac{e\,a}{m}\;.
\end{eqnarray}
For an electron or positron with momentum $p$ moving in an external
electromagnetic  field, its average effective momentum reads:
\begin{eqnarray}
q= p+\frac{e^2\,a^2\,k}{2k\cdot p}
 = p+\frac{\eta^2\,m^2}{2k\cdot p}\,k \; .
\end{eqnarray}
Here the second term is called as ponderomotive momentum. It is
proportional  to $a^2$ and hence represents a nonlinear effect. With
this momentum one may define an effective mass for the electron or
the positron by:
\begin{equation}
m_*^2=q^2=m^2+e^2a^2=(1+\eta^2)m^2\;.
\end{equation}
%

The wavefunction of the  electron $\Phi(x,p)$ or the positron
$\Phi'(x,p)$ is given by  the Volkov solution\cite{Volkov} of the
classic Dirac equation For circular polarization, it reads:
\begin{eqnarray}
 \Phi(x,p)&=& e^{iS_{-}[\psi]}\,\left(1-\frac{e\,/\!\!\!k \,/\!\!\!{A}}{2\,kp }\right)\;u(p)\nonumber\\
&\equiv &\Gamma(x,p)u(p) \; , \label{volkov-u}\\
 \Phi'(x,p)&=& e^{iS_{+}[\psi]}\,\left(1+\frac{e\,/\!\!\!k \,/\!\!\!{A}}{2(kp)}\right)\;v(p)\nonumber\\
&\equiv &\Gamma(x,p)v(p) \; , \label{volkov-v}
\end{eqnarray}
where
\begin{eqnarray}
S_{\pm}[\psi]&=&\pm(q x)+\frac{e(p
a_1)}{kp}\sin(kx)-\frac{e(p a_2)}{kp}\cos(kx) \\
\Gamma_{\pm}(x,p)&=&\sum_{s=-\infty}^{\infty}
 \exp[\pm i(q x-s k)x]\Big{[}B_s^0(\alpha_1^\pm, \alpha_2^\pm) \nonumber \\
  && +\frac{ea\, /\!\!\!k}{2k\cdot p}
 \left[\, /\!\!\!\epsilon_1 B_s^1(\alpha_1^\pm, \alpha_2^\pm)
 +\, /\!\!\!\epsilon_2 B_s^2(\alpha_1^\pm, \alpha_2^\pm)\right]\Big{]} \;,
\end{eqnarray}
where
\begin{eqnarray}
B_s^0(\alpha_1, \alpha_2)&=&\sum_{n=-\infty}^{+\infty}i^nJ_{s-n}(\alpha_1)J_n(\alpha_2)\nonumber\\
B_s^1(\alpha_1, \alpha_2)&=&\frac{1}{2}[B_{s-1}^0(\alpha_1, \alpha_2)+B_{s+1}^0(\alpha_1, \alpha_2)]\nonumber\\
B_s^2(\alpha_1, \alpha_2)&=&\frac{1}{2i}[B_{s-1}^0(\alpha_1,
\alpha_2)-B_{s+1}^0(\alpha_1, \alpha_2)]\nonumber \label{eq:bessel}
\end{eqnarray}

The generalized Bessel functions are given by sums over products of
the first kind.Similar treatment could be found in
\cite{bremsstrahlung}. The arguments of the generalized Bessel
functions are defined by $\alpha_1^\pm=\pm e\frac{p_\pm\cdot
a_1}{k\cdot p_\pm}$ and $\alpha_2^\pm=\mp e\frac{p_\pm\cdot
a_2}{k\cdot p_\pm}$. The electron spinors are used in the following
form:
\begin{eqnarray}
u_{ p}=\sqrt{\frac{E+m}{2E}} \left(
 \begin{array}{c}
   \zeta \\
   \frac{\mathbf{p}\cdot\mathbf{\sigma}}{E+m}\zeta\\
   \end{array}
 \right)
 \;\;\;
 v_{ p}=\sqrt{\frac{E+m}{2E}}
\left(
 \begin{array}{c}
   \frac{\mathbf{p}\cdot\mathbf{\sigma}}{E+m}\chi\\
    \chi\\
   \end{array}
 \right)
\end{eqnarray}
with the standard vector $\mathbf{\sigma}$ being composed of the
Pauli $2 \times 2$ spin matrices and $\zeta, \chi$ are two-component
Pauli spinors. $p$ is the momentum of the particle outside the laser
field.

The quantum field of  Dirac equation  can then be expressed as a
superposition of  the product of creation or annihilation operators
and Volkov solution at momentum $p$, instead of the plane wave
solutions for free particles.

The propagator of the electron in the laser background field can be
obtained by Fock-Schwinger proper time method\cite{zuber}. In a
circularly polarized plane it reads:
\begin{eqnarray}
S_A(x, x')=\left[\,i\, /\!\!\!\partial_x-e\,
/\!\!\!A(x)+m\,\right]\,(-i)\int_{-\infty}^{0}d\tau~ U(x, x';\tau)
\label{propagator}
\end{eqnarray}
with
\begin{eqnarray}
U(x, x';\tau)
 &=&-\frac{i}{(4\pi)^2\tau^2}
 \exp~i\Big{[}
 \frac{(x-x')^2}{4\tau}+\big(\,m_*^2+\frac{ea}{2}
 \phi^1_{\rho\nu}\sigma^{\rho\nu}
 \frac{\cos\xi_1-\cos\xi_2}{\xi_1-\xi_2}\nonumber\\
 &&+\frac{ea}{2}\phi^2_{\rho\nu}
 \sigma^{\rho\nu}\frac{\sin\xi_1-\sin\xi_2}{\xi_1-\xi_2}
 -i\epsilon\, \big)
 \tau-e\int^{x}_{x'}dy_{\mu}A^\mu(y)\Big{]}
\end{eqnarray}
with $\phi_{\rho\nu}=k_\mu\epsilon_\nu-k_\nu\epsilon_\mu$,
$\sigma^{\rho\nu}=\frac{i}{2}[\gamma_\mu, \gamma_\nu]$.
 Performing U's Fourier transform and integrating $\tau$, we can evaluate  the electron propagator.

With  Volkov solution given in (\ref{volkov-u}), (\ref{volkov-v})and
the propagator(\ref{propagator}),
the effects caused by the laser EM field can be explicitly extracted
out. Substituting  them into Eq~.(\ref{S-Psgg}), it follows  that:
\begin{eqnarray}
S_{Ps\to\gamma\gamma}&=&ie^2\,
  \; \int d^4x_1\, d^4x_2
 \;  e^{ i\,  k_1  x_1 + i\,k_2 x_2   } \nonumber \\
&&\int\frac{d^3 {\rm\bf q}}{(2\pi)^3}\tilde{\Phi}(\mathbf{p})Tr\{\hat{\Gamma}(x,p_+)(\, /\!\!\!\epsilon_1 S_A(x_1, x_2)\, /\!\!\!\epsilon_2\nonumber\\
&&+\, /\!\!\!\epsilon_2 S_A(x_1, x_2)\,
/\!\!\!\epsilon_1)\Gamma(x,p_-)\Pi_s\} \;.
 \label{eq:Sps}
\end{eqnarray}

and $\Phi(x,p)$, $\Phi'(x,p)$ are the Ps radial wave function in the
momentum space and projection operators which are given by:
\begin{eqnarray}
\Pi_0(P)&=& \frac{1}{\sqrt m}\,\gamma_5\, (\, /\!\!\!P +2m ) \;,
\end{eqnarray}
for the spin-singlet case
\begin{eqnarray}
\Pi_1(P ) &=&
 \frac{1}{\sqrt m}\, \not \!\!\epsilon \, (\, /\!\!\!P +2m )  \;,
\end{eqnarray}
for the spin-triplet state with polarization $\epsilon$.

We assume the ground state Ps atom  be initially at rest. The bound
state is a linear superposition of products of free states
$\Psi_{p_\pm}$ for the electron and positron with definite relative
momenta $\mathbf{q}$, which is weighted by the wavefunction
$\tilde{\Phi}(\mathbf{q})$. In the momentum space, the wavefunction
of the $S-$ wave reads:
\begin{eqnarray}
\tilde{\Phi}(\mathbf{q})=\frac{8\pi a_0^{3/2}}{[1+a_0^2
\mathbf{q}^2]^2} \;
\end{eqnarray}
where $a_0$ is the Ps radius.

To carry out integration over the relative momentum
$\tilde{\Phi}(\mathbf{q})$, we ignore $\tilde{\Phi}(\mathbf{q})$ in
those terms which are insensitive to $\mathbf{q}$ and keep those
terms which are sensitive to $\mathbf{q}$. The amplitude can be
expressed as the overlap integrals of the wave function and those
terms sensitive to $\mathbf{q}$ with the generic form:
\begin{eqnarray} \label{phi}
\phi=\sum_{r=-\infty}^{\infty}\int\frac{d^3{\rm\bf q}}{(2\pi\,
\sqrt{a_0})^3}\tilde{\Phi}(\mathbf{q})B_r^0(\alpha_+, \alpha_-) \;.
\end{eqnarray}
This means that the factorization here is different from the
conventional  factorization. Besides the wavefunction, there is
sensitive  $\mathbf{q}$ dependence arising from  Volkov state
wavefunction as expected. It is factorized in the time direction but
not in the space direction since they are in the same size in the
space direction. Actually the electron and the positron in  Volkov
states absorb or emit $s^-$ and $s^+$ laser photons respectively in
this process, corresponding to Bessel function$J_{s^-}(\alpha_-)$
and $J_{s^+}(\alpha_+)$. The Bessel functions decrease very quickly
if $\mid s^{\pm}\mid>\mid\alpha_\pm\mid$. In our case we find $\mid
s^{\pm}\omega\mid\ll\mid m\mid$, so we may take an approximation
\begin{eqnarray}
\delta(q_++q_--k_1-k_2+nk)\simeq\delta(q_++q_--k_1-k_2).
\end{eqnarray}
With this approximation, we can reduce the infinite sum over the
product of the Bessel function of the electron and the Bessel
function of the positron to a single Bessel function as shown in
Appendix. Finally, we can express it as
\begin{eqnarray}
\phi=\frac{16}{(2\pi a_0^{1/2})^3}\int_{0}^{+\infty}dq_z
\int_{0}^{+\infty}q_{\bot}dq_{\bot}\int_{0}^{\frac{\pi}{4}}d \varphi
\tilde{\Phi}(\mathbf{q})\exp{\frac{im\eta q_{\bot}\sin \varphi
}{\omega \sqrt{m^2+q_{\bot}^2+q_z^2}+q_z}}\;. \label{phi-int}
\end{eqnarray}
In the nonrelativistic limit, the factor in the exponential of the
integrant is simplified as $ i \frac{\eta}{\omega}\,
q_{\bot}\sin\varphi$. If the integration is from $0$ to $2\pi$, the
the overlap integral $\phi$ is nothing but Fourier transformation of
the wavefunction in momentum space, i.e., the wavefunction in
coordinator space at distance $\frac{\eta}{\omega} $ away from the
origin along the $q_{\bot}$ axes. For the Ps, it is proportional to
$ e^{- \frac{\eta}{2\omega a_B}}$ with $a_B$ being the Bohr radius.
This exponential suppression factor is in agreement with the result
given in \cite{Brazil} for the linear polarized laser, where the
authors obtained their results by taking a lot of approximations.
But for the exact expression given here, the integration bound is
from $0$ to $ \pi/4 $.

 From this expression, we can  evaluate
$\phi$ in (\ref{phi}) numerically. With this numerical value we
calculate the $S$-matrix element analytically.

The differential decay width can then be  expressed as
\begin{eqnarray}
d\Gamma=\frac{1}{16 \pi^2}\;|S_{Ps\rightarrow2\gamma}|^2\;
 \frac{d^3k_1}{2\omega_1}\,
 \frac{d^3k_2}{2\omega_2}
  \, \delta^{(4)}(q_++q_--k_1-k_2+nk)\;
\end{eqnarray}

We  carry out the phase space integration $\int\Pi$ by virtual of
the $\delta$-function
$\delta^4(q_++q_--k_1-k_2)=\delta(q_+^0+q_-^0-k_1^0-k_2^0)
\delta^3(\mathbf{q_+}+\mathbf{q_-}-\mathbf{k_1}-\mathbf{k_2})$ and
the relations $d^3k_2=\omega_2^2d\omega_2d\Omega$ . Integrating over
$d^3\mathbf{k_2}$ and $dk_1^0$ we find that the phase  space
integral can be expressed as
\begin{eqnarray}
\int\Pi&=&\frac{d^3k_1}{2\pi^3}\frac{1}{2\omega_1}
 \frac{d^3k_2}{2\pi^3}\frac{1}{2\omega_2}\\
&\approx&\frac{1}{16\pi^2}\int d
\Omega\frac{\omega_1}{2Eq+\frac{2e^2a^2}{Eq}(1-\cos\theta)}
\end{eqnarray}
where $\theta$is angle between photon $\gamma_1$ and the laser beam
direction, and
\begin{eqnarray}
\omega_1 &=& \frac{2E^3+2E\eta^2m^2}
{2E^2+\eta^2m^2-\eta^2m^2\cos\theta}\nonumber
\end{eqnarray}
%

Finally, we obtain the differential distribution over $\cos \theta
$.  For the spin-singlet state, it reads:
\begin{eqnarray}
\frac{d\Gamma^{cir}(\text{p-Ps})}{d\cos\theta}
 =\frac{\phi^2}{\pi^4}\Gamma_0(\text{p-Ps})
  \Big{[}1+\left(\frac{3\cos\theta}{2}-
  \frac{5}{2}\right)\eta^2\Big{]} \;,
\label{eq:p-Ps}
\end{eqnarray}
and for  the spin-triplet state, it reads:
\begin{eqnarray}
 \frac{d\Gamma_0^{cir}(\text{o-Ps})}{d\cos\theta}
 &=&
 \frac{\phi^2}{\pi^4}\Gamma_0(\text{p-Ps})
 \Big{(}\cos^2\theta+7\Big{)}\eta^2 \;.
 \label{eq:o-Ps}
\end{eqnarray}
Integrating out $\theta$ angle, one obtain the total decay rate.
\begin{eqnarray}
{d\Gamma^{cir}(\text{p-Ps})}
 &=& \frac{\phi^2}{\pi^4}\Gamma_0(\text{p-Ps})
  \Big{[}1 -
  \frac{5}{2}\eta^2\Big{]} \;,
\label{tot:p-Ps}
\end{eqnarray}
and for  the spin-triplet state, it reads:
\begin{eqnarray} \label{tot:o-Ps}
 \frac{d\Gamma_0^{cir}(\text{o-Ps})}{d\cos\theta}
 &=&
 \frac{\phi^2}{\pi^4}\Gamma_0(\text{p-Ps})
 \, 9 \eta^2 \;.
\end{eqnarray}
where $\Gamma_0(\text{p-Ps} \rightarrow\gamma\gamma)$ is the
two-photon decay width of the spin-single $1^1S_0$ Ps state in
vacuum. It reads:
\begin{eqnarray}
\Gamma_0(\text{p-Ps}\longrightarrow\gamma\gamma)=\frac{\alpha^5m_e}{2}\;.
\label{eq:freep-Ps}
\end{eqnarray}
Eqs.~(\ref{eq:p-Ps})-(\ref{tot:o-Ps}) are our final results. $\Gamma
$ from these equations we see that once we evaluate value of $\phi$,
we can make predictions for the lifetime of the Ps in the laser EM
field.

\section{RESULTS AND DISCUSSION}

In this section, we use Eqs.~(\ref{phi-int}),
(\ref{eq:p-Ps})-(\ref{tot:o-Ps}) derived in the last section to
calculate the decay rate of the Ps states in the strong laser EM
field. The only unknown parameter we need to calculate numerically
is the overlap integral $\phi$. As argued above, the electron and
the positron make period movement along the electric field provided
by the laser. This efficiently reduces the value of the wavefunction
at origin. As discussed  in the last section, to some extent, this
value is related to the wavefunction somewhere away from the origin
for the Ps with laser EM fields. In the positronium case, it is
exponentially suppressed comparing to the wavefunction at origin.
This exponential factor governs the enhancement of the lifetime of
the Ps states. The factor in the exponential is proportional to the
product of the wavelength squared and the square root of the
intensity of the laser field. Therefore, it is  more efficient to
enhance the lifetime of the positronium by increasing the wavelength
than increasing the intensity of the laser. For comparison with the
result   given in\cite{Brazil}, we first evaluate  Ti:sapphire
($f=380$ THz) laser source with intensity $10^{13}$ ${\rm W/cm^2 }$.
The result given in \cite{Brazil} is that the lifetime of the p-Ps
state is enhanced by $641 $ times while our result for this number
is $3.41 \times 10^3$. Their result is about 5 times smaller than
our one. Considering the fact that this enhancement is exponential
function and any tiny change of small factor will change this result
dramatically this difference is not very suppressing.

We now  turn to look at the case with the fixed  laser photon energy
at 1 eV but various  intensities. Our results are listed in I.

\begin{table}
\centering \caption{The enhanced factor for the lifetime of the p-Ps
and the o-Ps state in the laser field with 1 eV laser photon and
various value of $\eta $ } \setlength{\tabcolsep}{4pt}
\begin{tabular}{|c|c|c|c|c|}
\hline \hline
 ~ $\eta $~& $\Gamma_0(\text{p-Ps})/\Gamma_L(\text{p-Ps})$   & $\Gamma_0(\text{p-Ps})/\Gamma_L(\text{o-Ps})$\\
 \hline
 $0.01$& 54&$7.35\times10^{4}$
 \\
 \hline
 $0.05$& 454 &$2.5\times10^{4}$
 \\
 \hline
 $0.1$& 3630 &$5.1\times10^{4}$
 \\
 \hline

\end{tabular}
\end{table}

Finally we  turn to look at the case with the fixed  laser photon
energy with $10^{13}\; {\rm W/cm^2}$ intensity and  wavelength at
$1\, \mu m $  and $10 \; \mu m $ for the p-Ps state.  Our results
are listed in II. We see that for the wavelength at $10 \; \mu m $,
the lifetime of the $p-Ps$ state enhanced by a factor $10^8$.

\begin{table}
\centering \caption{ {The enhanced factor for the lifetime of the
p-Ps  states in the laser field with $10^{13}\; {\rm W/cm^2}$
intensity and  various wavelength }                }
\setlength{\tabcolsep}{4pt}
\begin{tabular}{|c|c|c|c|c|c|}
\hline \hline
   $\lambda(\mu m)$ ~&$ \Gamma_{0}$($/\Gamma{L}$) &Lifetime$(s)$\\
 \hline

 $1 $&$ 4.7 \times10^{3}$ &$5.96\times10^{-7}$
 \\
 \hline
  $10$&$ 1.3\times10^{8} $ &$1.5876\times10^{-2}$
 \\
 \hline

\end{tabular}
\end{table}



\section{Summary}
\label{sec:summary}

In this paper,  we calculate the partial lifetime of both the o-Ps
and the p-Ps positronium annihilation decay $ Ps\to\gamma\gamma$ in
the strong circular polarized laser EM field. We take a strategy of
the factorization to separate the effects caused by the Coulomb
interaction and the strong laser field. This factorization is
different from the conventional factorization. It is factorized in
the time direction but not in the space direction since they are at
the same size in the space direction. In calculating the
short-distance effects, we use the solution of  classic Dirac
equation in the laser EM background field, i,e., the so-called
Volkov state. We also have adopted Fock-Schwinger proper time
method. we expand the Volkov solutions in plane wave in terms of the
generalized Bessel function.  Our results show that for
long-wavelength laser with sufficiently high value of $\eta$
parameter which characterize the nonlinear effects of the laser
assisted process, the lifetimes of those Ps states are dramatically
increased. This is qualitatively consistent with that given in
\cite{Brazil}. This effect may be very useful for the HHG effects by
the Ps\cite{keitel2004}.


\begin{acknowledgments}
   This work is supported by the National Natural Science
Foundation of China under grants No.~112755242. We thank Xiao-Yuan
Li, Hai-Ting Chen, Gao-Liang Zhou for helpful discussions.
\end{acknowledgments}
\appendix
\section{GENERALIZED BESSEL FUNCTIONS}

Bessel functions of the first kind $J_n(x)$ , are solutions of
Bessel's differential equation. It is defined as

\begin{eqnarray}
  J_n(x)=\sum_{n=0}^{+\infty}\frac{(-1)^m}{m!\Gamma(m+n+1)}(\frac{1}{2}x)^{2m+n}
\end{eqnarray}

One important relation for integer orders is the Jacobi-Anger
expansion:
\begin{eqnarray}
e^{iz\cos\phi}&=&\sum_{n=-\infty}^{+\infty}i^nJ_n(z)e^{in\phi}\\
e^{iz\sin\phi}&=&\sum_{n=-\infty}^{+\infty}J_n(z)e^{in\phi}
\end{eqnarray}
With the help of Graf's addition theorem\cite{graf}, for $|ye^{\pm
i\phi}|<|x|$
\begin{eqnarray}
J_s(z)(\frac{x-ye^{-i\phi}}{x-ye^{i\phi}})^{\frac{s}{2}}=\sum_{n=-\infty}^{+\infty}J_{s+n}(x)J_n(y)e^{in\phi}
\end{eqnarray}
with $z=(x^2+y^2-2xy\cos\phi)^{1/2}$. So we have $B_s^0(\xi,
\eta)=J_{-s}(\sqrt{\xi^2+\eta^2})(\frac{-\xi-\eta i}{-\xi+\eta
i})^{s/2}$ for $|\eta|<|\xi|$, i.e.$|\sin\theta|<|\cos\theta|$. Note
that
\begin{eqnarray}
\frac{-\alpha_1-\alpha_2 i}{-\alpha_1+\alpha_2 i}&=&\frac{\alpha\cos\theta-\alpha\sin\theta i}{\alpha\cos\theta-\alpha\sin\theta i}\nonumber\\
&=&\frac{\cos\theta-\sin\theta
i}{\cos\theta+\sin\theta i}\nonumber\\
&=&(\cos\theta-\sin\theta i)^2
\end{eqnarray}
$\theta$ coincides with the azimuthal angle $\varphi$ for both
electron and positron. Then we use Graf's addition theorem again
\begin{eqnarray}
B_s^0(\alpha^+, \alpha^-)=J_{-s}(\alpha^\pm\cos\theta)\exp\frac{is
\pi}{4}
\end{eqnarray}
with $\alpha^\pm=\frac{eap_\bot}{k\cdot(\frac{P}{2}\pm q)}$ . For
integer order n, Bessel function $J_n$ is often defined via a
Laurent series for a generating function:
\begin{eqnarray}
\exp{\frac{x}{2}(t-\frac{1}{t})}=\sum_{n=-\infty}^{\infty}J_n(x)t^n,
\end{eqnarray}
\begin{eqnarray}
\sum_{n=-\infty}^{\infty}B_s^0(\alpha^+,
\alpha^-)=\exp{\frac{ieaq_\bot\sin \varphi}{k\cdot(\frac{P}{2}\pm
q)}}.
\end{eqnarray}
Thus we have,
\begin{eqnarray}
\phi=\frac{16}{(2\pi\,\sqrt{a_0})^3}\int_{0}^{+\infty}dq_z
\int_{0}^{+\infty}q_{\bot}dq_{\bot}\int_{0}^{\frac{\pi}{4}}d \varphi
\tilde{\Phi}(\mathbf{q})\exp{\frac{im\eta q_{\bot}\sin \varphi
}{\omega \sqrt{m^2+q_{\bot}^2+q_z^2}+q_z}}\;.
\end{eqnarray}
with
\begin{eqnarray}
\tilde{\Phi}(\mathbf{q})=\frac{8\sqrt{\pi}
a_0^{3/2}}{[1+(a_0q_\bot)^2+(a_0q_z)^2]^2}
\end{eqnarray}
and $a_0$ is the Bohr radius.We calculate this integral numerically.

\end{document}